\documentclass [a4paper,fleqn, 12pt]{article}
\usepackage{graphicx}

\usepackage {amsmath} \usepackage{amssymb} \usepackage{cite}
\newcommand{\cn}

\begin{document}

%\begin{frontmatter}

\title
{Nonlinear waves in bubbly liquids with consideration for viscosity and heat transfer}

\author
{ Nikolay A. Kudryashov,  \and Dmitry I. Sinelshchikov}

\date{Department of Applied Mathematics, National Research Nuclear University
MEPHI, 31 Kashirskoe Shosse,
115409 Moscow, Russian Federation}

%\author

%\corauthref{cor}},
%\corauth[cor]{Corresponding author.} \ead{nakudr@gmail.com}

%\address

\maketitle

\begin{abstract}

Nonlinear waves are studied in a mixture of liquid and gas bubbles. Influence of viscosity and heat transfer is taken into consideration on propagation of the pressure waves. Nonlinear evolution equations of the second and the third order for describing nonlinear waves in gas-liquid mixtures are derived. Exact solutions  of these nonlinear evolution equations are found. Properties of nonlinear waves in a liquid with gas bubbles are discussed.

\end{abstract}

%\begin{keyword}

% Point vortices, Special polynomials, Adler -- Moser polynomials

%\PACS 02.30.Hq - Ordinary differential equations

%\end{keyword}

%\end{frontmatter}

\section{Introduction}

A mixture of liquid and gas bubbles of the same size may be considered as an example of a classic nonlinear medium. In practice analysis of propagation of the pressure waves in a liquid with gas bubbles is important problem. Similar two - phase medium describes many processes in nature and engineering applications. In particular, such mathematical models are useful for studying dynamics of  contrast agents in the blood flow at ultrasonic researches \cite{Goldberg, Dayton}. The literature on this subject deals with theoretical and experimental studies of the various aspects for propagation of the pressure waves in bubbly liquids.

The first analysis of a problem bubble dynamics was made by Rayleigh \cite{Rayleigh}, who had solved the problem of the collapse of an empty cavity in a large mass of liquid. He also considered the problem of a gas - filled cavity under the assumption that the gas undergoes isothermal compression \cite{Plesset}. Based on the works of Rayleigh \cite{Rayleigh} and Foldy \cite{Foldy} van Wijngarden \cite{Wijng} showed that in the case of one spatial dimension, the propagation of linear acoustic waves in isothermal bubbly liquids, wherein the bubbles are of uniform radius, is described by the linear partial differential equation of the fourth order \cite{Jordan2006}.

The dynamic propagation of acoustic waves in a half - space filled with a viscous,  bubbly liquid under van Wijngaarden linear theory was considered in the recent work \cite{Jordan2006}. However we know that there are solitary and periodic  waves in a mixture of a liquid and gas bubbles and these waves can be described by nonlinear partial differential equations. As for examples of nonlinear differential equations to describe the pressure waves in bubbly liquids we can point out the Burgers equation \cite{Nakoryakov_1972, Bateman, Burgers}, the Korteweg -- de Vries equation \cite{Wijngaarden, Korteweg, Whitham, Drazin}, the Burgers -- Korteweg -- de Vries equation \cite{Wijngaarden} and so on.

Many authors applied the numerical methods to study properties of the nonlinear pressure waves in a mixture of liquid and gas bubbles. Nigmatullin and Khabeev \cite{Khabeev1974} studied the heat transfer between a gas bubble and a liquid by means of the numerical approach. Later Aidagulov and et al. \cite{Khabeev1977} investigated the structure of shock waves in a liquid with gas bubbles with consideration for the heat transfer between gas and liquid. Oganyan in \cite{Ogonyan00,Ogonyan05} tried to take into account the heat transfer between a gas bubble and a liquid to obtain the nonlinear evolution equations for the description of the pressure waves in a gas-liquid mixture. However, we have some different characteristic times of nonlinear waves in these processes and it turned out that the solution of this task is difficult problem.

The purpose of this work is to obtain more common nonlinear partial differential equations for describing the pressure waves in a mixture liquid and gas bubbles taking into consideration the viscosity of liquid and the heat transfer. We also look for exact solutions of these nonlinear differential equations to study the properties of nonlinear waves in a liquid with gas bubbles.

This Letter is organized as follows. System of equations for description of nonlinear waves in a mixture of liquid and gas bubbles taking into consideration for the heat transfer and the viscosity of liquid is given in section 2. In section 3 we obtain the basic nonlinear evolution equation to describe the pressure waves in liquid with gas bubbles. In sections 4 and 5 we present nonlinear evolution equations of the second and the third order and search for exact solutions of these nonlinear differential equations.

\section{System of equations for description of motion of liquid with gas bubbles with consideration for heat exchange and viscosity}

Suppose that a mixture of a liquid and gas bubbles is homogeneous medium \cite{Nakoryakov}. In this case for description of this mixture we use the averaged temperature, velocity, density and  pressure. We also assume that the gas bubbles has the same size and the amount of bubbles in the mass unit is constant $N$. We take the processes of the heat transfer and viscosity on the boundary of bubble and liquid into account. We do not consider the processes of formation, destruction, and conglutination for the gas bubbles.

We have that the volume and the mass of gas in the unit of the mass mixture can be written as
\begin{equation*}
V=\frac{4}{3}\pi R^{3}N,\,\,\,\,\,X=V\rho_{g},
\end{equation*}
where $R=R(x,t)$ is bubble radius, $\rho_{g}=\rho_{g}(x,t)$ is the gas density. Here and later we believe that the subscript $g$ corresponds to the gas phase and
subscript $l$ corresponds to the liquid phase.

We consider the long wavelength perturbations in a mixture of the liquid and the gas bubbles  assuming that characteristic length of waves of perturbation more than distance between bubbles.
We also assume, that distance between bubbles much more than the averaged radius of a bubble.

We describe dynamics of a bubble using the Rayleigh -- Lamb equation. We also take the equation of energy for a bubble and the state equation for the gas in a bubble into account. The system of equation for the description of the gas bubble takes the form \cite{Nakoryakov, Nigmatulin}
\begin{equation}
\rho_{l}\left(R R_{tt}+\frac{3}{2} R_{t}^{2}+\frac{4\nu}{3R}
R_{t}\right)=P_{g}-P, \label{eq: Relei}
\end{equation}

\begin{equation}
P_{g,\,t}+\frac{3nP_{g}}{R}R_{t}+\frac{3\chi_{g}\,Nu\,(n-1)}{2R^{2}}(T_{g}-T_{l})=0,
\label{eq: Energy}
\end{equation}

\begin{equation}
T_{g}=\frac{T_{0}P_{g}}{P_{g,\,0}}\left(\frac{R}{R_{0}}\right)^{3},
\label{eq: K-M}
\end{equation}
where  $P(x,t)$ is a pressure of a gas-liquid mixture, $P_{g}$ is a gas pressure in a bubble, $T_{g}$ and $T_{l}$ are temperatures of liquid and gas accordingly, $\chi_{g}$ is a coefficient of the gas thermal conduction, $Nu$ is the Nusselt number, $n$ is a polytropic exponent, $\nu$ is the viscosity of a liquid.

The expression for the density of a mixture can be presented in the form \cite{Nakoryakov}
\begin{equation}
\frac{1}{\rho}=\frac{1-X}{\rho_{l}}+V\,\Rightarrow \rho=\frac{\rho_{l}}{1-X+V\rho_{l}}.
\label{eq: density}
\end{equation}

Considering the small deviation of the bubble radius in comparison with the averaged radius of bubble,  we have
\begin{equation}
  \begin{gathered}
 R(x,t) = R_{0} + \eta(x,t),\:\   \quad
  R_{0}=const,\:\   \quad ||\eta||<<R_{0},\quad
  R(x,0)=R_{0}.
  \end{gathered}
  \label{eq: rel1}
\end{equation}

Assume that the liquid temperature is constant and equal to the initial value
\begin{equation*}
T_{l}=T\mid_{{t=0}}=T_{0},\,\,\,\,\,T_{0}=const. \label{eq: T_eq}
\end{equation*}

At the initial moment, we also have
\begin{equation*}
  \begin{gathered}
   t=0: \quad P=P_{g}=P_{0},\,\,\,\,\,P_{0}=const,
    \quad V=V_{0}=\frac{4}{3}\pi R_{0}^{3}N.
   \end{gathered}
\end{equation*}

Substituting $P_g$ and $T_g$ from Eqs.(\ref{eq: Relei}) and (\ref{eq: K-M}) into Eq.(\ref{eq:
Energy}) and taking relation (\ref{eq: rel1}) into account we have the pressure dependence of a mixture on the radius perturbation in the form
\begin{equation}
  \begin{gathered}
P-P_{0}+\frac{\eta}{R_{0}}P+\frac{3n\varkappa}{R_{0}}P
\eta_{t}+\varkappa P_{t}+
\frac{\rho_{l}(3R_{0}^{2}+4\nu\varkappa)}{3R_{0}}\eta_{tt}+
\frac{\rho_{l}(6R_{0}^{2}-4\nu\varkappa)}{3R_{0}^{2}}\eta
\eta_{tt}+\\
+\frac{\rho_{l}(8\nu\varkappa(3n-1)+9R_{0}^{2})}{6R_{0}^{2}}\eta_{t}^{2}+
\frac{4\nu\rho_{l}}{3R_{0}}\eta_{t}+
\frac{2P_{0}}{R_{0}}\eta
+\frac{ 3 P_{0}}{R_{0}^{2}}\eta^{2}=0,\vspace{0.2cm}\\
\\
\varkappa=\frac{2R_{0}^{2}P_{0}}{3\chi Nu (n-1) T_{0}}.
    \end{gathered}
  \label{eq: state_eq}
\end{equation}

From Eq.(\ref{eq: density}) we also have the dependence $\rho$ on $\eta$ using formula (\ref{eq: rel1})
\begin{equation}
\begin{gathered}
\rho=\rho_{0}-\mu \eta +\mu_{1} \eta^{2}, \quad
\rho_{0}=\frac{\rho_{l}}{1-X+V_{0}\rho_{l}},\\
\\
\mu=\frac{3\rho_{l}^{2}V_{0}}{R_{0}(1-X+V_{0}\rho_{l})^{2}}, \quad
\mu_{1}=\frac{6 \rho_{l}^{2} V_{0}(2 \rho_{l}
V_{0}-1+X)}{R_{0}^{2}(1-X+\rho_{l} V_{0})^{3}}.
 \label{eq: desity1}
\end{gathered}
\end{equation}

We use the system of equations for description of the motion of a gas-liquid mixture flow  in the form
\begin{equation}
  \begin{gathered}
  \frac{\partial \rho}{\partial t} + \frac{\partial (\rho\,u)}{\partial x} = 0,  \quad
  \rho\left(\frac{\partial u}{\partial t} + u\,\frac{\partial u}{\partial x}\right)
  +  \frac{\partial P}{\partial x}=0,   \hfill
  \end{gathered}
  \label{eq: continuity_and_E_eq}
\end{equation}
where $u=u(x,t)$ is a velocity of a flow of a gas-liquid mixture.

Eq.(\ref{eq: continuity_and_E_eq}) together with Eqs.(\ref{eq:
state_eq}) and \eqref{eq: desity1} can be applied for description of nonlinear waves in a gas-liquid medium.

Consider the linear case of the system of equations   \eqref{eq: state_eq},  \eqref{eq: desity1}and \eqref{eq: continuity_and_E_eq}. Assuming, that pressure in a mixture is proportional to perturbation radius,  we obtain the linear wave equation for the radius perturbations
\begin{equation}
\eta_{tt}=c_{0}^{2} \, \eta_{xx}, \quad c_{0}=\sqrt{\frac{3P_{0} }{\mu R_{0}}}.
\end{equation}

Let us introduce the following dimensionless variables
\begin{equation*}
\begin{gathered}
  t = \frac{ l }{ c_{0} }\, t', \quad   x = l\, x', \quad  u = c_{0}\, u', \quad
  \eta=R_{0} \eta^{'}, \quad P = P_0\, P'+P_0,
  \label{eq: non-dim_subst}
\end{gathered}
\end{equation*}
where $l$ is the characteristic length of wave.

Using the dimensionless variables the system of equations  \eqref{eq: state_eq},  \eqref{eq: desity1} and \eqref{eq: continuity_and_E_eq} can be reduced  to the following (the primes of the variables are omitted)
\begin{equation}
  \begin{gathered}
  \eta_{t}-\frac{\rho_{0}}{\mu R_{0}}u_{x} + u \eta_{x}+\eta u_{x}-\frac{2\mu_{1} R_{0}}{\mu} \eta \eta_{t}= 0,  \hfill \\
  \\
  -\frac{\rho_{0}}{\mu R_{0}}\left(u_{t}+u u_{x}\right)+\eta u_{t}- \frac{1}{3} P_{x}=0,   \hfill \\
  \\
  P+\varkappa_{1} P_{t}+\eta P +3n\varkappa_{1}\,\eta_{t} P=-
  (\beta_{1}+\beta_{2})\,\eta_{tt}-  (2\beta_{2}-\beta_{1})\, \eta\eta_{tt}-\hfill \\-\left(\frac{3n-1}{2}\beta_{1}+\frac{3}{2}\beta_{2}\right)\eta_{t}^{2}- \lambda \eta_{t}-3 \eta+3\eta^{2},  \hfill
  \end{gathered}
  \label{eq: main non-dim system}
\end{equation}
where the parameters are determined by formulae
\begin{equation}
  \begin{gathered}
    \lambda = \frac{ 4 \nu \rho_{l}c_{0} }{ 3 P_0\, l }+3n\varkappa_{1},\quad
     \beta_{1}=\frac{4\nu\,\varkappa\,\rho_{l}\, c_{0}^2}{3\,P_{0}\,l^{2}},\quad
    \beta_{2} = \frac{\rho_{l}\, c_{0}^2\,R_{0}^2}{P_{0}\,l^{2}},\\
    \gamma=\frac{\varkappa\rho_{l} R_{0}^{2}c_{0}^{3}}{P_{0}l^{3}},\,\quad
    \varkappa_{1}=\frac{\varkappa c_{0}}{l}.
  \end{gathered}
  \label{eq: non-dim_parameters}
\end{equation}

\section{Basic nonlinear evolution equation for description of nonlinear waves in liquid with bubbles}

We can not find the exact solutions of the system of nonlinear differential equations (\ref{eq: main non-dim system}). To account for the slow variation of the waveform, we introduce a scale transformation of independent variables  \cite{Kudryashov}
\begin{equation}
\begin{gathered}
  \xi = \varepsilon^m(x-t), \quad \tau= \varepsilon^{m+1}\, t ,\quad
  m > 0, \quad \varepsilon \ll 1,
  \label{eq: rescaling_coordinates}
    \end{gathered}
\end{equation}

\begin{equation*}
  \frac{\partial}{\partial x} = \varepsilon^m \frac{\partial}{\partial \xi},\quad
  \frac{\partial}{\partial t} = \varepsilon^{m+1} \frac{\partial}{\partial \tau}
  - \varepsilon^m \frac{\partial}{\partial \xi}.
\end{equation*}

Substituting (\ref{eq: rescaling_coordinates}) into (\ref{eq: main
non-dim system}) and dividing  on $\varepsilon^{m}$  in first two equations we have the following system of equations

\begin{equation}
  \begin{gathered}
    \varepsilon \eta_{\tau} - \eta_{\xi} -\frac{\rho_{0}}{\mu R_{0}} u_{\xi}
      +\, \eta u_{\xi}
      +\, u \eta_{\xi} -\varepsilon \frac{2\mu_{1} R_{0}}{\mu} \eta \eta_{\tau}
      + \, \frac{2\mu_{1} R_{0}}{\mu} \eta \eta_{\xi} = 0,
\end{gathered}
  \label{eq: general_rescaled_system1}
\end{equation}

\begin{equation}
  \begin{gathered}
    \varepsilon \left(-\frac{\rho_{0}}{\mu R_{0}}\right) u_{\tau} +\frac{\rho_{0}}{\mu R_{0}} u_{\xi}
    +\varepsilon \eta u_{\tau} - \, \eta u_{\xi}
      - \, \frac{\rho_{0}}{\mu R_{0}} u u_{\xi}
      - \frac{1}{3}\,  P_{\xi} = 0,
\end{gathered}
  \label{eq: general_rescaled_system2}
\end{equation}

\begin{equation}
  \begin{gathered}
    P +\varepsilon^{m+1} \varkappa_{1} P_{\tau}-\varepsilon^{m} \varkappa_{1} P_{\xi}
     +\, \eta P+\varepsilon^{m+1} 3n\varkappa_{1} \eta_{\tau} P-\varepsilon^{m} 3n\varkappa_{1} \eta_{\xi} P=\hfill \vspace{0.1cm} \\=
      -\varepsilon^{2m+2} \,(\beta_{1}+\beta_{2})\, \eta_{\tau\tau}\,
      + \, \varepsilon^{2m+1} 2 \,(\beta_{1}+\beta_{2})\, \eta_{\tau\xi}\,
      - \, \varepsilon^{2m} \,(\beta_{1}+\beta_{2}) \, \eta_{\xi\xi}\,-\hfill \vspace{0.2cm} \\
      -\varepsilon^{2m+2} \,(2\beta_{2}-\beta_{1})\, \eta \eta_{\tau\tau}\,+
      \, \varepsilon^{2m+1} 2 \,(2\beta_{2}-\beta_{1})\, \eta \eta_{\tau\xi}\,-\hfill \vspace{0.2cm} \\
      - \, \varepsilon^{2m} \,(2\beta_{2}-\beta_{1}) \, \eta \eta_{\xi\xi}\,
      -\varepsilon^{2m+2}\left(\frac{3n-1}{2} \beta_{1}+\frac{3}{2}\beta_{2}\right)\eta_{\tau}^{2}
      +\hfill \vspace{0.3cm} \\+\varepsilon^{2m+1} 2\left(\frac{3n-1}{2} \beta_{1}+\frac{3}{2}\beta_{2}\right)\eta_{\tau}\eta_{\xi} -
     \varepsilon^{2m}\left(\frac{3n-1}{2} \beta_{1}+\frac{3}{2}\beta_{2}\right) \eta_{\xi}^{2}
      %-\hfill\\
       -\hfill \vspace{0.3cm} \\-\, \varepsilon^{m+1} \lambda \, \eta_{\tau}\,
      + \,   \varepsilon^{m}  \lambda\, \eta_{\xi}-3 \, \eta\, +\,3 \eta^{2}. \hfill
      %-\,4 \eta^{3}
  \end{gathered}
  \label{eq: general_rescaled_system3}
\end{equation}

We now assume that the state variables $u$, $\eta$ and $P$ can be represented asymptotically as series in powers of $\varepsilon$ about an equilibrium state
\begin{equation}
  %\left\{
  \begin{gathered}
    u = \varepsilon u_1 + \varepsilon^{2} u_2 +  \ldots,  \,\,\,
    \eta  =\varepsilon \eta _1 + \varepsilon^{2} \eta _2 +\ldots,  \,\,\,
    P = \varepsilon P_1  + \varepsilon^{2} P_2   +\ldots
  \end{gathered}
  %\right.
  \label{eq: asymptotic_expansion1}
\end{equation}

Substituting (\ref{eq: asymptotic_expansion1}) into (\ref{eq: general_rescaled_system1})-(\ref{eq: general_rescaled_system3}) and equating expressions  at  $\varepsilon$ and $\varepsilon^2$ to zero, we obtain some equations with respect to $u_1$, $u_2$, $\eta_1$, $\eta_2$, $P_1$ and $P_2$. Solving these equations with respect to $P_1$ we have  the equation for pressure $P_1$ in the form
\begin{equation}
\begin{gathered}
 P_{1\tau}\,+\alpha\,P_{1}P_{1 \xi}+
 \varepsilon^{2m-1}\, \frac{\beta_{1}+\beta_{2}}{6}\, P_{1\xi\xi\xi}\,
-\varepsilon^{2m}\, \frac{2\beta_{2}-\beta_{1}}{18}\, P_{1}
P_{1\xi\xi\xi}\,- \hfill \\- \varepsilon^{2m}\,
\left(\frac{3n-2}{18}\,\beta_{1}+\frac{5}{18}\,\beta_{2}\right)\,
P_{1\xi} P_{1\xi\xi}\, =
\varepsilon^{m-1} \, \left(\frac{\lambda}{6}\, - \, \frac{\varkappa_{1}}{2}\right) \,
P_{1\xi\xi} +\hfill \vspace{0.1cm} \\
+\varepsilon^{m} \, \frac{n\varkappa_{1}}{2}
\,\left(P_{1}P_{1\xi}\right)_{\xi},
 \label{eq:master equation1}
  \end{gathered}
\end{equation}
where
\begin{equation}
\begin{gathered}
 \alpha=\frac{3\mu R_{0}}{ \rho_{0}}-\frac{3\mu_{1}R_{0}}{\mu}+\frac{2}{3}.
 \label{eq:master equation1a}
  \end{gathered}
\end{equation}

From Eq. \eqref{eq:master equation1} one can find some nonlinear evolution equations for description of waves in a mixture liquid and gas bubbles with consideration for the heat transfer and the viscosity.

\section{Nonlinear evolution equation of second order for description of waves in liquid with bubbles}

Assuming $m=1$,  $(\beta_{1}+\beta_{2})/6=\beta\,\varepsilon^{\delta}$,  $\delta>0$, $\beta\sim1$  in Eq.\eqref{eq:master equation1}, we have  nonlinear evolution equation of the second order in the form
\begin{equation}
P_{1 \tau}+\alpha \,P_{1}P_{1\xi}=\left(\frac{\lambda}{6}-\frac{\varkappa_{1}}{2}\right)
P_{1 \xi\xi}+\varepsilon \, \frac{n\varkappa_{1}}{2} \,
\left(P_{1} P_{1\xi}\right)_{\xi}.
\label{eq: extBurgers_equation}
\end{equation}

Using transformations  $P_{1}=1/\varepsilon\, P_{1}^{'}$ and $\alpha=\varepsilon \, \alpha^{'}$ one can write Eq.\eqref{eq: extBurgers_equation} in the form
\begin{equation}
P^{'}_{1 \tau}+\alpha^{'} P_{1}^{'}P^{'}_{1\xi}=\Lambda
P^{'}_{1 \xi\xi}+ \Lambda_{1} \,
\left(P^{'}_{1} P^{'}_{1\xi}\right)_{\xi},\,
\label{eq: extBurgers_equation1}
\end{equation}

\begin{equation*}
\Lambda=\frac{\lambda}{6}-\frac{\varkappa_{1}}{2}=
\frac{c_{0}}{3\,l}\left(\frac{2\nu\rho_{l}}{3P_{0}}-\frac{P_{0}R_{0}^{2}}{\chi\,
Nu\,T_{0}}\right),\,
\label{c1}
\end{equation*}

\begin{equation*}
\Lambda_{1}=\frac{n\varkappa_{1}}{2}=\frac{n\,c_{0}\,P_{0}\,R_{0}^{2}}{3\,\chi\,
Nu\,(n-1)\,T_{0}\,l}.
\label{c11}
\end{equation*}

We can see that dissipation of the nonlinear waves is described by the equation \eqref{eq: extBurgers_equation1} and depends on values of coefficients $\Lambda$ and $\Lambda_{1}$.

The first term in expression for $\Lambda$ corresponds to a dissipation of a wave because of the viscosity on boundary a bubble-liquid. The second term in $\Lambda$ corresponds to the heat transfer between a liquid and bubbles. We have to note that there is the decrease of the coefficient $\Lambda$ in the case of the heat transfer. The coefficient $\Lambda_{1}$  characterizes the nonlinear dissipation of a wave which is explained by the heat transfer between liquid and gas bubbles.

In the case of the isothermal process ($Nu\rightarrow\infty,\,\varkappa_{1}\rightarrow0$) Eq.\eqref{eq: extBurgers_equation1} goes to the Burgers equation. We have the dissipation of the wave that is determined by the value of the viscosity of a liquid. Currently the Burgers equation is well known. The wave processes described by this equation are well known. Taking the Cole -- Hopf transformation into account this equation can be reduced to the heat equation [11, 12]. Analytical solutions of many problems described by this equation were constructed.

The nonlinear term in a right hand side \eqref{eq: extBurgers_equation1} leads to the additional damping of the pressure wave in comparison with the Burgers equation. These nonlinear dissipative effects are connected with the heat transfer in a mixture liquid and gas bubbles.

One can construct some exact solutions of the equation \eqref{eq: extBurgers_equation1}. Using the scale transformations in Eq. \eqref{eq: extBurgers_equation1}
\begin{equation}
\xi=\frac{n\,\varkappa_{1}}{2\alpha'}\,\xi^{'},\quad
\tau=\frac{6}{(\lambda-3\,\varkappa_{1})}\,\left(\frac{n\,\varkappa_{1}}{2\alpha'}
\right)^{2}\tau^{'},\quad
P^{'}_{1}=\frac{(\,\lambda-3\,\varkappa_{1})}{3\,n\,\varkappa_{1}}\, v,
\label{eq: coor_stretch_B}
\end{equation}
we obtain the following equation (the primes are omitted)
\begin{equation}
v_{\tau}+ v\,v_{\xi}=v_{\xi\xi}+\left(v\,v_{\xi}\right)_{\xi}.
\label{eq: extBurgers_equation1_sl_t}
\end{equation}

Some exact solutions of Eq.\eqref{eq: extBurgers_equation1_sl_t} can be found.
Taking transformation \eqref{eq: coor_stretch_B} into account  we obtain the solution of Eq. \eqref{eq: extBurgers_equation1} by the formula
\begin{equation}
\begin{gathered}
P_{1}^{'(1,2)}(\xi,\tau)=\frac{(\,\lambda-3\,\varkappa_{1})}{3\,n\,\varkappa_{1}} \left[\left( \frac{3}{4\left(  C_{0}+1 \right)} \pm \frac {3}{\sqrt {16\, (C_{0}+1)^{2}+9\,C_{2}\,{\rm e}^{-\frac{\theta}{2}}}}\right) ^{-1}- 1\,\right],\\
\\
\theta=\frac{2\alpha'}{n\,\varkappa_{1}}\xi-C_{0}\,\frac{(\lambda-3\,\varkappa_{1})}{6}\,
\left(\frac{2\alpha'}{n\,\varkappa_{1}}\right)^{2}\tau.
\label{eq: extBurgers_sl}
\end{gathered}
\end{equation}

From expression \eqref{eq: extBurgers_sl} one can see, that the amplitude of a wave depends on its velocity. Besides the amplitude of this pressure wave is  defined by a relation between viscosity of a liquid and a gas thermal conduction.

Dependencies of solution $P_{1}^{'(1)}$   from $\xi$  at different values parameter  $\varkappa_{1}$ at $C_{0}=2, C_{2}=1, \tau=0$  are illustrated on Fig.1. We use values $\varkappa_{1}$ are following: 0.200; 0.205; 0.210.  From Fig.1 we can see that  the amplitude of a wave decreases with growth  $\varkappa_{1}$. Similar profiles of the pressure waves were observed experimentally. They are called by shock waves with monotone structure or waves \cite{Wijngaarden1974}.

\begin{figure}[h]
\begin{center}
 \includegraphics[width=90mm,height=60mm]{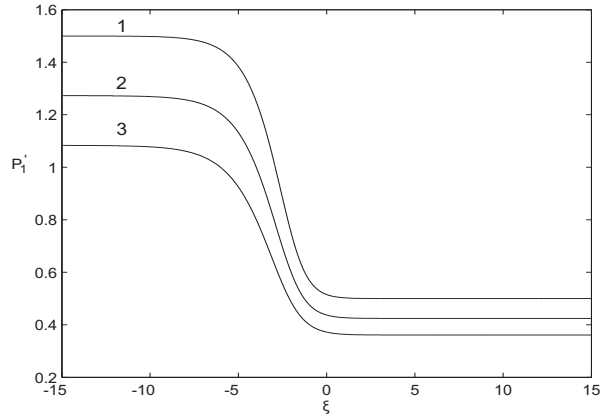}
 \caption{Kink wave solution \eqref{eq: extBurgers_sl} of Eq. \eqref{eq: extBurgers_equation1}  at $\varkappa=0,20;\,0,22;\,0,24$ (curves 1, 2, 3)}
\end{center}
 \label{fig1}
\end{figure}

Experimental study of the damping of the pressure waves in a liquid with a mixture of bubbles of two different gases were presented in \cite{Nakoryakov2002a}. In particular it was found the damping of amplitude for the pressure wave for a liquid with freon bubbles. The volume gas content of freon bubbles  was $\alpha=1.2\%$ and the diameter of bubbles was $d=2,3$ mm.

We have considered the propagation of the damping of the pressure impulse which is described by Eq.\eqref{eq: extBurgers_equation1}. For numerical modeling of the solitary waves we use the difference equation which corresponds to Eq.\eqref{eq: extBurgers_equation1}.   We have observed the good agreement of results of numerical modeling with experimental data. This comparison is given on Fig. 2.

\begin{figure}[h]
\begin{center}
 \includegraphics[width=95mm,height=65mm]{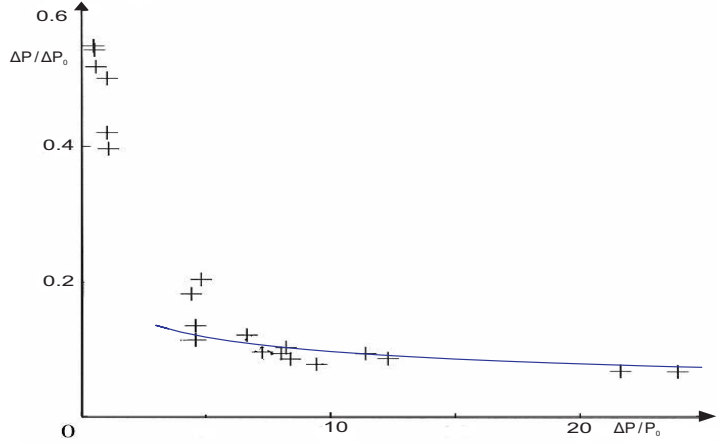}
  \caption{Comparison of amplitude attenuation  of a pressure wave described by the equation \eqref{eq: extBurgers_equation1} with experimental data}
\end{center}
 \label{fig2}
\end{figure}

Let us discuss the features of the wave dynamics described by the equation \eqref{eq: extBurgers_equation1}. For numerical modeling of the solitary waves we use the difference equation which corresponds to Eq.\eqref{eq: extBurgers_equation1}.

Comparison of the solitary pressure pulses described by the Eq.\eqref{eq: extBurgers_equation1} with the solitary pressure pulses of the the Burgers equation (($\Lambda_{1}=0$ in \eqref{eq: extBurgers_equation1})is given on the Fig. 3. We can see from Fig. 3  that pressure front described by Eq.\eqref{eq: extBurgers_equation1} is more smooth than pressure front described by Burgers equation. So the heat transfer between the gas in bubbles and the liquid leads to additional dissipation of the solitary wave.

\begin{figure}[h] %[!hp]
  \centering      %width=7cm,height=7cm
 \includegraphics[width=0.49\textwidth]{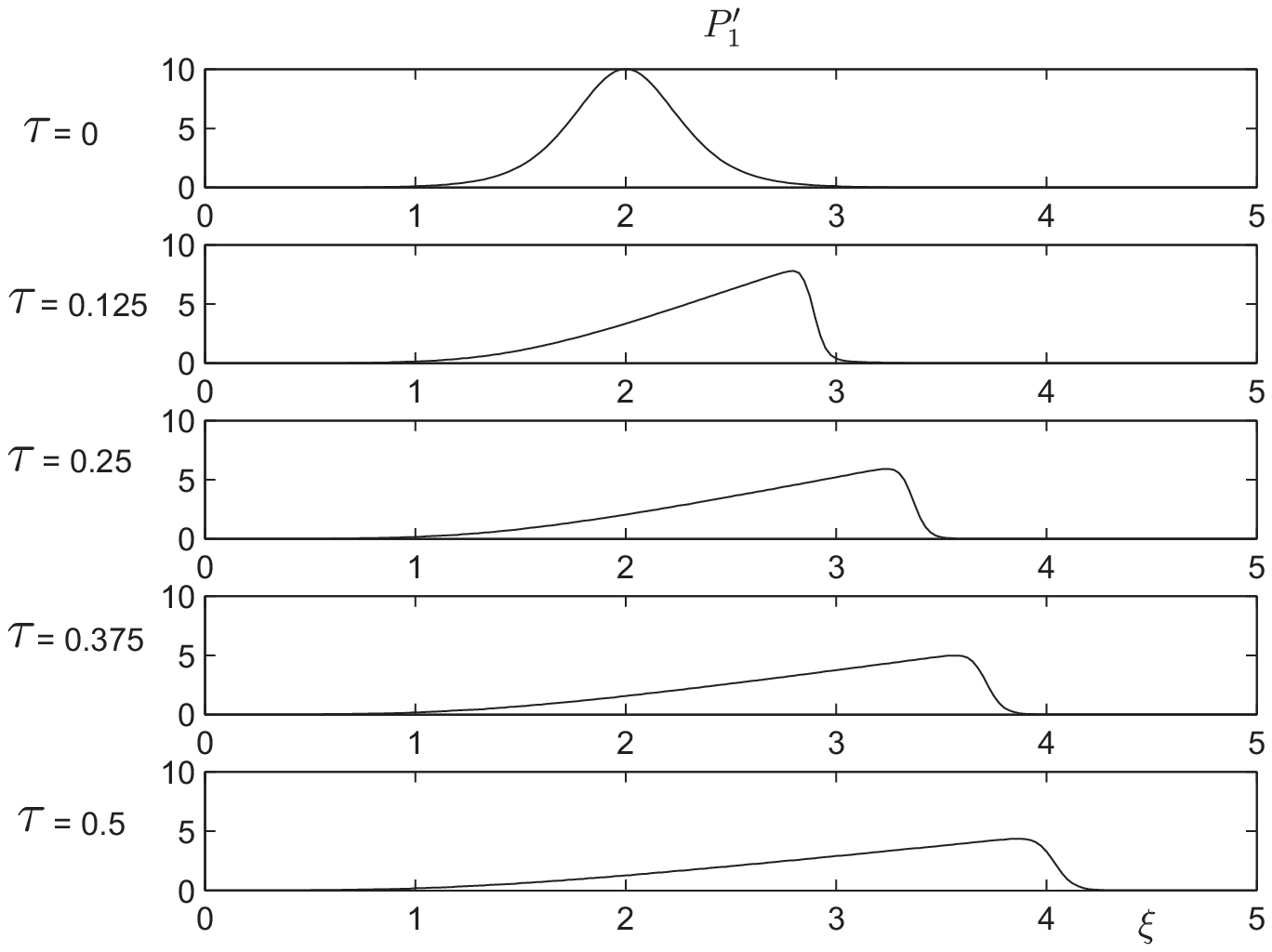}
 \includegraphics[width=0.49\textwidth]{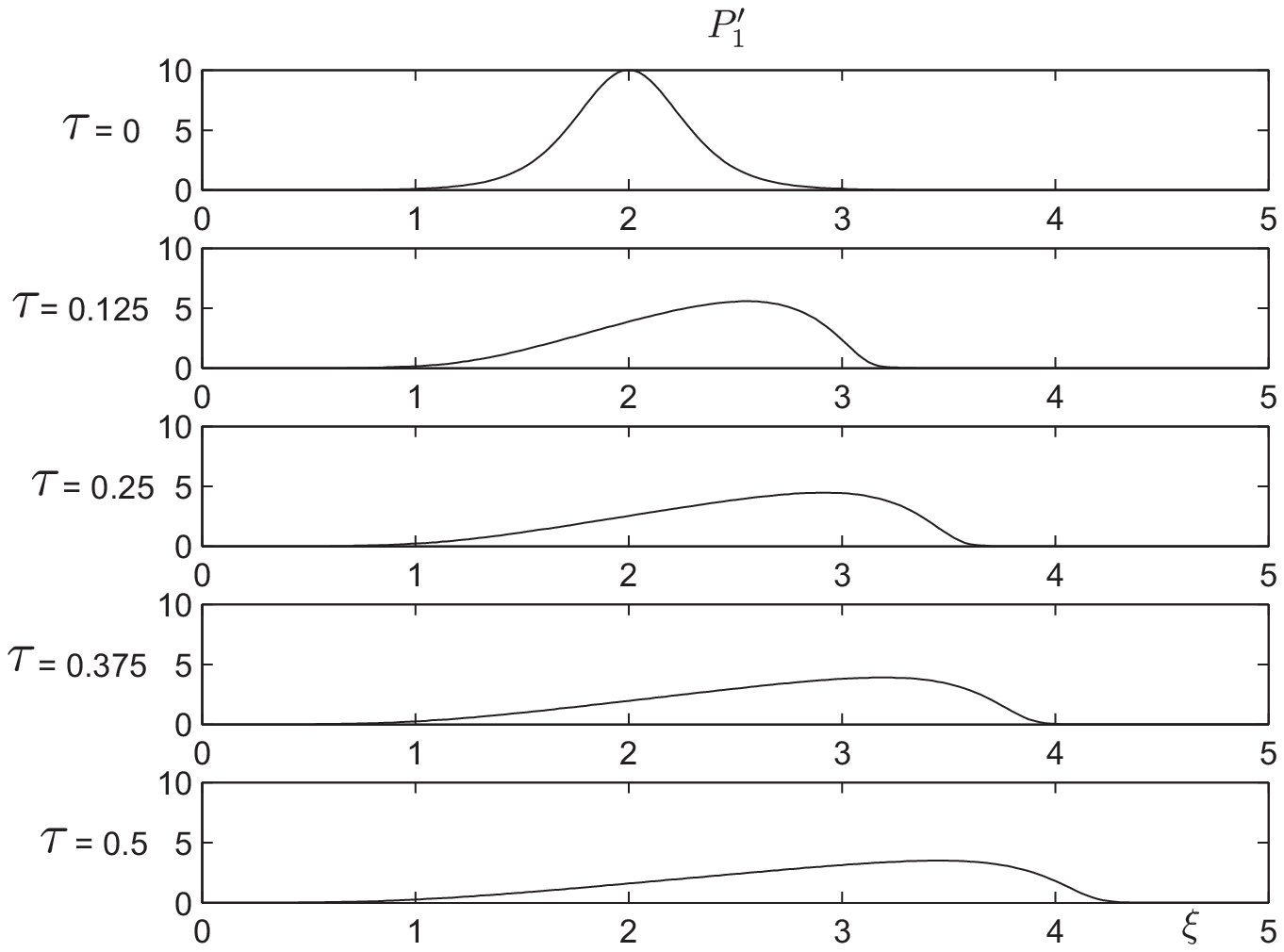}
    \caption{Evolution of the pressure solitary wave described by Burgers equation (left figure)  and Eq. \eqref{eq: extBurgers_equation1} (right figure).  }
  \label{fig:solitary}
\end{figure}

\section{Nonlinear evolution equation of the third order for description of waves in a liquid with gas bubbles}

Assuming that $m=\frac{1}{2}$,  ${\lambda}=6\,\varepsilon^{1/2}\,\lambda^{'}$,
 $\lambda^{'}\sim1$,  ${\varkappa_{1}}={2}\,\varepsilon^{1/2}\,
 \varkappa_{1}^{'}$, $\varkappa_{1}^{'}\sim1$ in Eq.(\ref{eq:master equation1}) and using transformations $P_{1}=\dfrac{1}{\varepsilon}\, P_{1}^{'}$, $\alpha=\varepsilon \, \alpha^{'}$ we have the nonlinear evolution equation of the third order
\begin{equation}
\begin{gathered}
P^{'}_{1\tau}+\alpha^{'} P^{'}_{1}P^{'}_{1\xi}+\frac{\beta_{1}+\beta_{2}}{6}\,
P^{'}_{1\xi\xi\xi}-  \frac{2\beta_{2}-\beta_{1}}{18}\,
P^{'}_{1}P^{'}_{1\xi\xi\xi}\,-\hfill \\-\left(\frac{3n-2}{18}\,\beta_{1}+\frac{5}{18}\,\beta_{2}\right)\,
P^{'}_{1\xi} P^{'}_{1\xi\xi} =\left(\lambda^{'}-\varkappa_{1}^{'}\right)
P^{'}_{1 \xi\xi}+ n\varkappa_{1}^{'}  \,
\left(P^{'}_{1} P^{'}_{1\xi}\right)_{\xi}.
\label{eq: extBKdV1}
\end{gathered}
\end{equation}

The coefficient of dispersion can be determined taking  the relations \eqref{eq: non-dim_parameters} into account
\begin{equation}
B=\frac{\beta_{1}+\beta_{2}}{6}=
\frac{R_{0}^{2}\,c_{0}^{2}\,\rho_{l}}{6\,l^{2}\,P_{0}}\left(1+\frac{8\nu\,P_{0}}{9
\chi_{g}\,Nu\,(n-1)T_{0}}\right).
\label{c2}
\end{equation}

We can see that the value of dispersion depends on the initial radius of a bubble and a relation between a viscosity and the coefficient of the  heat transfer on interphase boundary. In the case of the isothermal condition for gas in bubbles ($Nu\rightarrow\infty$) or in the case of small viscosity of a liquid ($\nu \rightarrow 0$) the wave dispersion is defined by value of radius of bubbles.

Taking $Nu\rightarrow\infty$ and $\nu \rightarrow 0$ into account in Eq.\eqref{eq: extBKdV1} we have Korteweg -- de Vries equation that was obtained for description of nonlinear waves in a liquid with gas bubbles \cite{Wijngaarden}.  Using $Nu\rightarrow\infty$  we have from Eq.\eqref{eq: extBKdV1} the Burgers -- Korteweg -- de Vries equation, obtained in \cite{Nakoryakov_1972}. This equation is not integrable but some special solutions of Burgers -- Korteweg -- de Vries equation were obtained in \cite{Kudryashov88, Kudryashov07a}. It is likely that the nonlinear evolution equation \eqref{eq: extBKdV1} is new for the description of nonlinear waves in a liquid with gas bubbles.

Consider Eq.\eqref{eq: extBKdV1} without a dissipation. In this case we have
\begin{equation}
\begin{gathered}
P^{'}_{1\tau}+\alpha P^{'}_{1}P^{'}_{1\xi}+\frac{\beta_{1}+\beta_{2}}{6}\,
P^{'}_{1\xi\xi\xi}-  \frac{2\beta_{2}-\beta_{1}}{18}\,
P^{'}_{1}P^{'}_{1\xi\xi\xi}\,-\\-\left(\frac{3n-2}{18}\,\beta_{1}+\frac{5}{18}\,\beta_{2}\right)\,
P^{'}_{1\xi} P^{'}_{1\xi\xi} =0.
\end{gathered}
\label{eq: ext_KdV_sl}
\end{equation}
Using the scale transformations
\begin{equation}
\tau=\frac{6}{\beta_{1}+\beta_{2}}\,\tau^{'},\quad P^{'}_{1}=\frac{3(\beta_{1}+\beta_{2})}{2\beta_{2}-\beta_{1}}\,v,
\label{eq: stretching_transform}
\end{equation}
we obtain equation (the primes are omitted)
\begin{equation}
v_{\tau}+\alpha_{1} v\,v_{\xi}+\,
v_{\xi\xi\xi} =  \,(v\,v_{\xi\xi})_{\xi}+\beta\,v_{\xi} v_{\xi\xi},
\label{eq: ext_KdV_sl_2}
\end{equation}
\begin{equation*}
\begin{gathered}
\alpha_{1}=\frac{9\alpha'}{2\beta_{2}-\beta_{1}}=\frac{81\alpha'P_{0}l^{2}\chi\,Nu(n-1)\,T_{0}}
{2\rho_{l}\,c_{0}^{2}\,R_{0}^{2}(9\chi Nu (n-1)T_{0}-4\nu P_{0})}, \vspace{0.1cm} \\
\\
\beta=\frac{(3n-1)\beta_{1}+3\beta_{2}}{2\beta_{2}-\beta_{1}}=\frac{27\chi\,Nu (n-1)T_{0}+8\,Nu(3n-1)P_{0}}{9\chi\,Nu(n-1)T_{0}-4\,\nu\,P_{0}}.
\end{gathered}
\end{equation*}

The traveling wave reduction of Eq.\eqref{eq: ext_KdV_sl_2} coincides with the Camassa - Holm equation \cite{Camasa, Parkes} that has one of interesting class of solutions in the form of peaked  solitons . The most simple form of these solutions is described by the formula
\begin{equation}
v(\xi,\tau)=A\,\exp(-|k\,\xi-k^{3}\,\tau+B|),
\label{eq: peaked}
\end{equation}
where $k$ is wave number, $A$ and $B$ are arbitrary constants.
Taking
\begin{equation}
\beta=\frac{\alpha_{1}}{k^{2}}-2,
\label{eq: peaked_c}
\end{equation}
and $v_{\xi'\xi'\xi'}=-v_{\xi'\xi'\tau'}$ into consideration Eq.\eqref{eq: ext_KdV_sl_2} can be written as
\begin{equation}
m_{\tau'}+v\,m_{\xi'}+\left(\frac{\alpha_{1}}{k^{2}}-1\right)\,v_{\xi'}\,m=0,
\label{eq: peaked_eq}
\end{equation}
\begin{equation}
\begin{gathered}
v(\xi')=\frac{1}{2}  \int_{-\infty}^{+\infty} e^{-|\xi'-y|}\,m(y)\,dy,\quad m=v-v_{\xi'\xi'}, \quad \tau'=k^{3}\,\tau,\quad \xi'=k \xi.
\end{gathered}
\label{eq: peaked_eq1}
\end{equation}

Using relations  \eqref{eq: stretching_transform}, \eqref{eq: peaked_c}, we have the solution of Eq.\eqref{eq: ext_KdV_sl} in the form of the peaked solutions
\begin{equation}
P_{1}'(\xi,\tau)=\frac{3(\beta_{1}+\beta_{2})\,A }{2\beta_{2}-\beta_{1}} \,\exp(-|k\,\xi- \frac{(\beta_{1}+\beta_{2})\,k^{3}}{6}\tau+B|).
\label{eq: peaked_sl}
\end{equation}
The graph of a solution \eqref{eq: peaked_sl} at
$A=1/6,\,B=5,\,\beta_{2}=\beta_{1}=1,\tau=0$ is demonstrated on Fig.4.
\begin{figure}[h]
\begin{center}
\includegraphics[width=80mm,height=60mm]{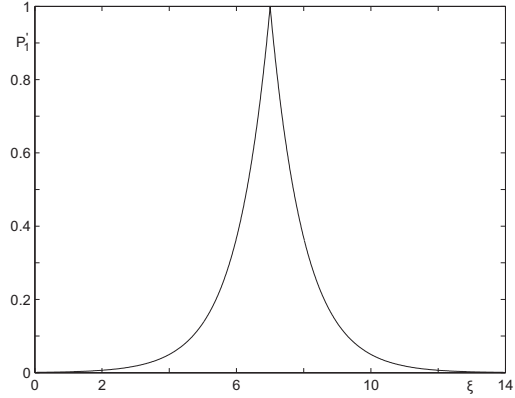}
  \caption{Peaked solution \eqref{eq: peaked_sl} of Eq. \eqref{eq: ext_KdV_sl}}
\end{center}
 \label{fig3}
\end{figure}

Using the traveling wave ansatz $v(\xi,\tau)=y(z), z=\xi-C_{0}\tau$ in Eq.(\ref{eq: ext_KdV_sl_2})  and integrating on $z$, we obtain the equation
\begin{equation}
C_{1}-C_{0}\,y+\frac{\alpha_{1}}{2}\,y^{2}+y_{zz}-y\,y_{zz}-\frac{\beta}{2} y^{2}_{z}=0.
\label{eq: r ext_KdV_sl_3}
\end{equation}

The general solution of Eq.\eqref{eq: r ext_KdV_sl_3} at
$C_{0}=\alpha_{1},\quad C_{1}=\alpha_{1}/2$
is expressed by formulae
\begin{equation}
\begin{gathered}
y(z)=1-\left(C_{3}\,e^{\int f(z)\,dz}\right)^{-1},\quad C_{3}\neq0,\,\quad f_{z}=\frac{\beta+2}{2}\,f^{2}-\frac{\alpha_{1}}{2},
\label{eq: r ext_KdV_solve}
\end{gathered}
\end{equation}
where $C_{3}\neq0$ is constant of integration.
Accordingly to \eqref{eq: r ext_KdV_solve} there are two families of solutions of Eq.\eqref{eq: r ext_KdV_sl_3}
\begin{equation}
y(z)=1-\frac{1}{C_{3}}\,\left[\textrm{tg}^{2}\left\{\frac{\sqrt{|\alpha_{1}(\beta+2)
|}}{2}\left(z+C_{2}\right)\right\}+1\right]^{-1/(\beta+2)},\quad \alpha_{1}(\beta+2)<0,
\label{eq: r ext_KdV_solution_1}
\end{equation}

\begin{equation}
y(z)=1-\frac{1}{C_{3}}\,\left[\tanh^{2}\left\{\frac{\sqrt{\alpha_{1}(\beta+2)}}{2}\left
(z+C_{2}\right)\right\}-1\right]^{-1/(\beta+2)},\quad \alpha_{1}(\beta+2)>0,
\label{eq: r ext_KdV_solution_2}
\end{equation}
where $C_{2}$ is arbitrary constant.

Family of solutions $\alpha_{1}(\beta+2)<0$ describes the periodic waves. In the case $\beta<-2$ solution \eqref{eq: r ext_KdV_solution_1} has the singular points that are determined by formula
\begin{equation*}
z_k=-C_{2}+\frac{2}{\sqrt{|\alpha_{1}(\beta+2)|}}\left(\frac{\pi}{2}+\pi k\right),\,k\in\mathbb{Z}.
\end{equation*}

Family solutions at $\alpha_{1}(\beta+2)>0$ and $\beta<-2$ are solitary waves. In the case  $\beta>-2$ solution goes to indefinite when $z\rightarrow \pm \infty$.
Dependence of this solutions from $z$  at parameter $\beta=-1$  at $C_{2}=0,\,C_{3}=1$  is presented on Fig. 5.

\begin{figure}[h]
\begin{center}
 \includegraphics[width=80mm,height=60mm]{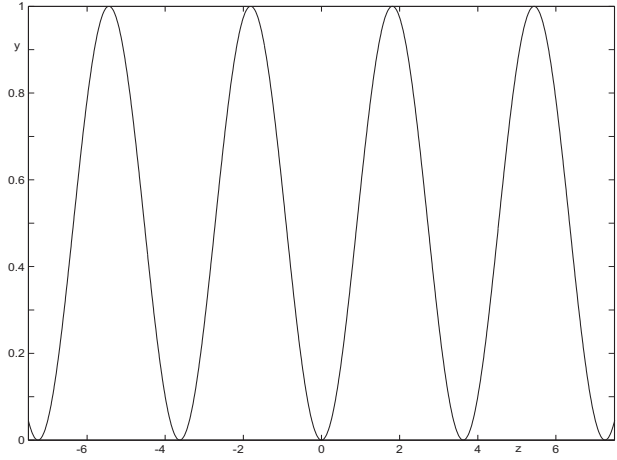}
   \caption{Periodical solution \eqref{eq: r ext_KdV_solution_1}  of Eq. \eqref{eq: r ext_KdV_sl_3} at $\beta=-1$}
\end{center}
 \label{fig4}
\end{figure}

One can find some exact solutions of the equation \eqref{eq: extBKdV1} with the dissipation. Using the scale transformations
\begin{equation}
\xi=\frac{2 \beta_{2}-\beta_{1}}{18\,n\,\varkappa^{'}_{2}}\,\xi^{'},\quad
\tau=\frac{6}{\beta_{1}+\beta_{2}}\,\left(\frac{2\beta_{2}-\beta_{1}}{18n
\varkappa^{'}_{1}}\right)^{3}\,\tau^{'},\quad
P^{'}_{1}=\frac{3(\beta_{1}+\beta_{2})}{2\beta_{2}-\beta_{1}}\,v,
\end{equation}
we obtain from equation \eqref{eq: extBKdV1} the following equation (the primes are omitted)
\begin{equation}
\begin{gathered}
v_{\tau}+\alpha_{1} \, v\,v_{\xi}+v_{\xi\xi\xi}
= \left(v\, v_{\xi\xi}\right)_{\xi}\, +
\beta \, v_{\xi}\, v_{\xi\xi}+\sigma \, v_{\xi\xi}+ \,
\left(v \,v_{\xi} \right)_{\xi},\,
\label{eq: extBKdV_sl_1}
\end{gathered}
\end{equation}
where
\begin{equation*}
\begin{gathered}
\alpha_{1}=\frac{(2\beta_{2}-\beta_{1})\alpha'}{18\,n^{2}\,\varkappa^{'2}_{1}},\quad
\beta=\frac{(3n-1)\beta_{1}+3\beta_{2}}{2\beta_{2}-\beta_{1}},\quad
\sigma=\frac{(\lambda^{'}-\varkappa'_{1})(2\beta_{2}-\beta_{1})}{3\,n\,\varkappa^{'}_{2}
(\beta_{1}+\beta_{2})}.
\end{gathered}
\end{equation*}

Eq.\eqref{eq: extBKdV_sl_1} does not pass the Painlev\'e test and is not integrable by the inverse scattering transform but there are some exact solutions of this equation. These solutions can be found using the simplest equation method \cite{Kudryashov_1990, Kudryashov_2005, Kudryashov07, Vitanov08} for the travelling wave $v(\xi,\tau)=y(z), z=\xi-C_{0}\tau$. We look for solution in the form
\begin{equation}
y(z)=a_{0}+a_{1}\,w(z)+a_{2}\,w(z)^{2},
\label{extBKdV_tr}
\end{equation}
where $w(z)$ is solution of linear ordinary differential equation of the second order \cite{Kudryashov_2008, Kudryashov_2009}
\begin{equation}
w_{zz}=A\,w_{z}+B\,w+C.
\label{eq:_extBKdV_SE}
\end{equation}

One of these solution at $\alpha_{1}<-1$ takes the form
\begin{equation}
\begin{gathered}
y(z)=a_{0}+\frac{\alpha_{1}(a_{0}-1)\,F}{2\,C}\left(1+\frac{\alpha_{1}}{8\,C}\,F\right), \vspace{0.3cm} \\
F=\left( C_{3}  \sin \left\{ \frac{\sqrt {-1-\alpha_{1}}z}{2} \right\}
+C_{2} \cos \left\{ \frac{\sqrt {-1-\alpha_{1}}z}{2} \right\} \right)e^{-\frac{z}{2}} -\,\frac {4C}{\alpha_{1}}.
\label{eq: sl_eBKdV}
\end{gathered}
\end{equation}
where $a_0$, $C$, $C_{2}$ and $C_{3}$ are arbitrary constants.

Exact solution \eqref{eq: sl_eBKdV} of Eq.\eqref{eq: extBKdV_sl_1} represents oscillating fast damping waves which can be observed at propagation of nonlinear waves in a mixture liquid and gas bubbles. Analyzing the wave dynamics described by  the nonlinear ordinary differential equation of the third order \eqref{eq: extBKdV_sl_1} we can see that this equation describes the different forms of the solitary and periodical waves.

\section{Conclusion}

We have considered nonlinear waves in a mixture of a liquid and gas bubbles taking into consideration the viscosity of liquid and the heat transfer between liquid and gas bubbles. We have applied different scales of time and coordinate. As a result we have obtained the nonlinear evolution equations \eqref{eq: extBurgers_equation1} and \eqref{eq: extBKdV1} of the second and the third order for describing the pressure waves in a liquid with bubbles. It is likely that these differential equations are new and generalize the Burgers equation, the Korteweg -- de Vries equation and the Burgers -- Korteweg -- de Vries equation.  These evolution equations allow us to take into account the influence of the viscosity and the heat transfer on the boundary of liquid and bubbles. Generally these equations are not integrable but all these equations have some exact solutions that have been found in this Letter.

We had observed that nonlinear evolution equation \eqref{eq: ext_KdV_sl} can be used for describing waves at characteristic time $t \sim \varepsilon^{-3/2}$. Without dissipative processes Eq.\eqref{eq: ext_KdV_sl} has a solution in the form of peaked solitons  (so-called peakons). Taking into account dissipative processes  we have solution of Eq.\eqref{eq: extBKdV1} for description of nonlinear waves. Exact solutions have been found and we obtain the damping of the periodic pressure waves.  To describe the pressure waves in the case of  $t \sim \varepsilon^{-2}$ the nonlinear evolution equation of the second order can be used. Using the travelling wave ansatz the analytical solutions of this equation have been obtained. These solutions correspond to shock waves with monotone structure.

\end{document}